\documentstyle[11pt]{article}
\def\be{\begin{equation}}
\def\ee{\end{equation}}
\def\bea{\begin{eqnarray}}
\def\eea{\end{eqnarray}}
\def\la{\label}
\def\ci{\cite}
\def\al{\alpha}
\def\O{\Omega}
\def\de{\delta}
\def\pp{\phi}
\def\no{\nonumber}
\def\le{\left}
\def\ri{\right}

\begin{document}

\begin{flushright}
hep-ph/9501250\\
IFUNAM-FT-94-64\\
\end{flushright}
\vspace{7 mm}

 \begin{center}
 {\LARGE \bf  Vanishing of the Cosmological  Constant, Stability of
the Dilaton
and Inflation} \\
 \vspace{1cm}
{\large  A. de la Macorra\footnote{E-mail address:
macorra@teorica0.ifisicacu.unam.mx}}\\
\vspace{3mm}
{\em Instituto de Fisica UNAM, Apdo. Postal  20-364,\\
01000 Mexico D.F., Mexico.
}\\ [8mm]
 \end{center}
\vspace{1cm}

\begin{abstract}
\noindent
We study the possibility of canceling the cosmological constant in
supergravity string models. We show that  with a suitable choice of
superpotential  the vacuum energy  may vanish with  the  dilaton
field at its
minimum and  supersymmetry broken    with a large hierarchy.  We
derive the
condition for which the introduction of a chiral potential, e.g. the
inflaton
potential,   does not  destabilize the dilaton field even  in the
region where
the   scalar  potential  takes positive values. This  allows  for an
inflationary potential with the dilaton frozen at its minimum. \\

\noindent
\rule[.1in]{12.5cm}{.002in}

\end{abstract}

\thispagestyle{empty}

\setcounter{page}{0}
\vfill\eject

\newpage

The vanishing of the cosmological constant remains an open issue
\ci{V=0}.
  In the absence of an  understanding of why  the  cosmological
constant
cancels one can study, nevertheless,  the  possibility of having a
potential
that  vanishes at the  global minimum.
In non-supersymmetric models one has, in principle, a large  freedom
to choose
arbitrary
terms and fine tune them to get zero cosmological constant. However,
in a
supersymmetric theory the potential  terms are  more constraint. In
fact, for
global supersymmetry (SUSY)
this is not possible if one requires  SUSY to be spontaneously
broken. On the
other hand, in supergravity
models one can have   the potential $V = 0$ and SUSY spontaneously
broken. The breaking of SUSY is certainly a necessary condition but
usually spontaneously broken symmetries have a potential
with  $V|_{min}=-\Lambda^4$,  where $\Lambda$ is the symmetry
breaking scale
and for a realistic  hierarchy solution  $V\simeq
-[10^{-6}\,m_{Planck}]^{4}$,
many orders of magnitude larger than the observational  upper limit
$|V| <
10^{-120} \,m_{Planck}^4$.
It is therefore interesting to determine
whether one can  have  vanishing cosmological constant with SUSY
broken  at a
phenomenological acceptable scale in supergravity models.

The cosmological constant  also plays an important role in the
evolution of the
universe.
Since string models \ci{strings} are valid below the Planck scale
they  should
describe  the evolution of the universe.  The standard big bang
theory  has
some shortcomings like the isotropy and flatness problems
\ci{faults}. An
inflationary epoch \ci{chao}, where the universe expanded in an
accelerated
way,  may solve these problems.
This inflationary  period  is also welcome to erase the abundance  of
the
topological defects produced when a symmetry is spontaneously broken.
 In order for the potential to inflate a necessary condition is to
have a
positive  energy.  For arbitrary values of the different fields one
expects V
to be positive and to evolve to its minimum. In this evolution one
would hope
for an inflationary period.
However,  it is
difficult to obtain an inflationary potential in string models due to
the
dynamics  of the dilaton field $S$ \ci{stein},\ci{infaxel}.
When  the scalar potential  $V$ evolves to the minimum of the
dilaton,  the
universe, keeping all  other fields fixed, does not  go trough an
inflationary
period.
At the minimum, SUSY is broken and for vanishing v.e.v. of the chiral
fields,
the vacuum energy is negative and of the order of $\Lambda^4$.
So, it seems that either we have a positive potential with the wrong
dynamics
due to dilaton  field
or when the dilaton is at its minimum we have  a  negative
potential.
We will show  that it  is indeed  possible to
have a positive inflationary potential and $S$ frozen.

The effective $ D=4 $
superstring  model
is given by an $N = 1$
supergravity theory \ci{sugra} with at least  two gauge singlet
fields
$S$ and $T$ as well as an
unspecified number of gauge
chiral matter superffields $\pp$. We will consider only an overall
moduli
$T$.
The v.e.v. of the dilaton field $S$ gives
the gauge
coupling constant $g^{-2} = {\rm Re} S$ at the string scale  while
the real
part of the moduli fields $Re\, T= R^2_i$ the
radius of the
compactified dimension.
The tree level scalar potential
is given by
\cite{sugra}
\be
V_{0} = \frac{1}{4} e^{K}  (G_{a} (K^{-1})^{a}_{b} G^{b} -3|W|^{2})
\label{eq1}\ee
where the Kahler potential is $G_{a}  \equiv K_{a}W + W_{a}$ and
\bea
K & = & - \log(S_{r}) -3
\log (T_{r} ) +T_r^n |\phi|^2
\label{eq2}\\
W & = &  W_0(S,T) + W_{ch}(T,\phi).
\nonumber\end{eqnarray}
The indices
$a,b$ run
over all chiral fields, i.e. the dilaton $S$, the moduli $T$ and
chiral fields
$\phi_i$ with modular weight $n$ and $T_{r} =   T + \bar{T}, S_{r}  =
S +
\bar{S} $.  In orbifold compactification $n$ can vary from 5 to -9
\ci{Tdual}.
All the fields  are
normalized with respect to the reduced Planck mass
 $m_{p} =  M_{p}/\sqrt{8 \pi}$. As usual, the indices $a,b$
of the functions $G, K$ and $W$ denote derivatives with respect to
chiral fields.

The superpotential term $W_0(S,T)$ arises due to non-perturbative
effects, like gaugino condensation \cite{gaug}, and is responsible
for
breaking supersymmetry (SUSY) while $W_{ch}$ is the chiral matter
superpotential.
  Another important property of string models is  duality invariance.
Under
this symmetry
 the moduli field $T$ transform as element of $SL(2,Z)$ group,  the
dilaton $S$
is invariant\footnote{If loop corrections are included there may be a
mixing
term between $S$ and $T$ in the Kahler potential $K$ and $S$ would no
longer be
invariant. However, this loop correction is expected to be small with
respect
to the tree level value and we will  for simplicity not consider it.}
while the
chiral fields $\phi$ transform as modular function with weight $n$.
The
superpotential must be a  modular function of weight -3. Using the
duality
symmetry one can then  fix the $T$ dependent part of  the
superpotential $W$ in
terms of the Dedekind-eta function $\eta$,
\be
W = \eta(T)^{-6} \Omega(S,\pp), \;\;\;\;\eta \equiv e^{-\pi
T/12}\Pi_n
(1-e^{-n\,\pi T})
\la{eq3}\ee
where we have  assumed, for simplicity,  that the $T$ dependent part
in $W$ can
be factorized and
\be
\O=\O_0(S)+\O_{ch}(\pp).
\la{eq4}\ee
$\O_0$ corresponds to the non-perturbative superpotential due to
gaugino
condensate while $\O_{ch}$  to the chiral matter superpotential.
Using eqs.(\ref{eq1}) and (\ref{eq4}) the scalar potential becomes
\be
V_0=  \frac{1}{4} e^{K}|\eta|^{-12} \left[
|h|^2+|k_i|^2+|\O|^2(\frac{3
T_r^2}{4\pi^2} |\hat{G}_2(T)|^2-3)\right]
\la{eq5}\ee
where $\hat{G}_2$ is the Eisenstein function of modular weight 2 and
we have
defined the auxiliary fields of $S$ and $\pp$ in terms of
$h=\eta^{6}\, S_r\,
G_S=S_r \O_S-\O$ and $k_i=\eta^{6} \,G_i=K_i \O+\O_i$ respectively
while the
auxiliary field of the moduli  $T$ is $G_T=\frac{3}{2\pi}
\eta^{-6}\,\O\, \hat
G_2(T)$.  Note that we have  taken  the chiral fields in
eq.(\ref{eq3}) to be
canonically normalized, i.e. $K_{\pp}^{\pp}= 1$.

 In the context of supergravity (tree level and one-loop level
potential
only),  the cancelation  of the cosmological constant must come
trough a
non-vanishing  auxiliary field $G_i \neq 0$.
In string models there is a large number of chiral matter fields
$\pp$ with
different interacting superpotential terms .
Since we don not wish to specify to any given model we will take
$\O_{ch}$ to
be arbitrary and we will study if there exist a particular
superpotential for
which SUSY is broken   with the dilaton at is minimum and   zero
cosmological
constant.

The condition of  zero cosmological constant, considering only the
tree level
potential, is
\be
G_{a} (K^{-1})^{a}_{b} G^{b} =3|W|^{2}
\la{eqq1}\ee
and it is hard to satisfy dynamically in the context of string
models.
In fact,  the stabilization of the dilaton field  is not  simple to
achieve and
there are two different approaches that give a large
hierarchy\footnote{We will
not consider here  the $S$-dual symmetry \ci{Sdual}}.  The first case
is to
consider two different gaugino condensate \ci{2gaug} with slightly
different
one-loop $\beta$-function coefficients and chiral matter  fields with
non-vanishing v.e.v.
In this case a stable solution is found for $G_S=0$.
Another possibility is to
 include one-loop  corrections of the 4-Gaugino interaction  and to
minimize
$V_0+V_1$ \ci{1gaug}, where $V_1$ is the one-loop potential. In this
case a
stable minimum is found for a single gaugino condensate. In either
case the
 effective $S$-dependent superpotential generated by a gaugino
condensate is
$\O_0(S)=d \,e^{-3 \,S/2b_0}$ with $b_0$ the one-loop beta function
coefficient
and $d$  independent of $S$.

\noindent
I) Two Gaugino Condensates

Let us  first consider the case of two gaugino condensates.  To find
the vacuum
state with zero cosmological constant  one needs to solve the eqs.
\be
V|=V_S|=V_T|=V_i|=0.
\la{eq6}\ee
The extremum eq. for the moduli field $T$ is
\be
V_T=(K_T- 6 \frac{\eta_T}{\eta})V+\frac{\partial}{\partial T} \,
\frac{1}{4}e^{K}|\O|^2\left[\frac{3 T_r^2}{4\pi^2}
|\hat{G}_2(T)|^2\right]=0
\la{eq6a}\ee
and using $V|=0$, eq.(\ref{eq6a}) is satisfied for $T$ at the dual
invariant
points ($T=1,e^{-\pi/6}$) where $\hat{G}_2=0$. This implies that the
auxiliary
field of the moduli is  zero,  $G_T=0$, and it does not break SUSY
contrary to
the case where the condition $V|=0$ was not imposed.  For $T$ at the
dual
points the vanishing of $V$ implies then that either $h$ or  $k$ or
both must
be different than zero.

The extremum eq. for  the dilaton field  is
 \be
V_S= \frac{1}{4}e^{K}|\eta|^{-12} \le[ h_S \bar h +h \bar h_S +k_S
\bar k
+k\bar k_S -\frac{1}{S_r}\le(|h|^2 + |k|^2 -3\,h\, \bar{\O}\ri)  \ri]
=0
\la{eq8}\ee
which implies that
\be
h_S \bar h +h \bar h_S - \frac{h}{S_r}(\bar h - 3\bar{\O})= -
\frac{\bar
k}{S_r} ( S_r k_S -k )
\la{eq7}\ee
where we have already taken $\bar k_S=0$.
{}From eq.(\ref{eq7}) we see that $h=0$ is valid only if  $\bar k (
S_r k_S -k
)=0$.  In \ci{2gaug} it was shown that for $k=0$, $h=0$ corresponds
to a
minimum while  $h \neq 0$ to a maximum. However, imposing the
condition $V|=0$
we can no longer  have $k=h=0$.  Assuming then that $k\neq 0$,
$V_S=h=0$ would
require
\be
S_r k_S-k=K_i (S_r \O_S-\O)+ S_r\O_{Si}-\O_i=0.
\la{eq9}\ee
The first term of eq.(\ref{eq9}) vanishes by hypothesis ($h=0$) and
taking
$\O_{si}=0$
 implies that $\O_i=0$.  Since
 $|k|^2=|K_i\O+\O_i|^2=|K_i\O|^2$ must cancel the term $3 | \O|^2$ to
have zero
cosmological constant,
a solution to $V|=V_S|=h=0$ is valid  only if
\be
\O_i=0\;\;\;\; {\rm and} \;\;\;\; K_i=  \bar{\pp_i} = \sqrt{3}.
\la{eq10}\ee
However, it is clear  that  this value cannot be dynamically obtained
since
the direction
$\pp \rightarrow 0$  gives $k=\bar \pp \O \rightarrow 0$ and  a
negative $V $.
We come to the conclusion that we cannot have $V|=V_S|=0$ and $h=0$.
A
different solution to $S$ must be found in the context of two gaugino
condensates.  But,  as we will show below, the new minimum for $S$
requiring
that $V|=0$ is not far away from the one
obtained for $h=0$ with $V| <0$.

Can we actually have $k\neq 0$ ? If all superpotential terms are at
least
quadratic in $\pp_i$ then $k=0$
for $\pp_i=0$. The only possibility  to have $k\neq 0$ is with a
linear
superpotential  $\O_{ch} =c \phi$, where $c$ is an arbitrary
constant.  Let us
take then the superpotential
\be
\O=\O_0(S)+c\;\pp.
\la{eq11}\ee
The auxiliary field $k=\bar \pp (\O_0+c\,\pp) +c \;\;\;$ may vanish
only if
\be
|\O_0|^2-4|c|^2 \ge 0.
\la{}\ee
 Therefore,  for  $c$   large enough  $k$ will be different  than
zero.
The scalar potential which is generated  by the superpotential
(\ref{eq11}) is
\be
V_0=  \frac{1}{4}e^{K}|\eta|^{-12} \left[ |S_r \O_S -(\O_0 +
c\,\pp)|^2+ |\bar
\pp (\O_0+c\,\pp) +c |^2 -3 |\O_0+c\,\pp|^2\right].
\la{eq12}\ee
In the case of two gaugino condensates $\O_0$ is
\be
\O_0= d_1\,e^{-3S/2b_1}+d_2\,e^{-3S/2b_2}
\la{eq13}\ee
where $b_i=\frac{3N_i-M_i}{16\pi^2}$ is the one-loop beta function
coefficient
of the i-th gauge groups and
$d_i=\le(\frac{M_i}{3}-N_i\ri)\le(32\pi^2\,e\ri)^{\frac{3(M_i-N_i)}{3N
_i-M_i}}\le(\frac{M_i}{3}\ri)^{\frac{M_i}{3N_i-M_i}}$ \ci{2gaug}.
The eqs. $V|=V_S|=V_{\pp}=0$ are
\bea
|h|^2+|k|^2 &=&3|\O|^2
\no\\
h_S \bar h +h \bar h_S &=& \O_S \le(3\bar \O  - \bar \pp \bar k \ri)
\no\\
k_{\pp} \bar k +k \bar k_{\pp} &=& \O_{\pp} \le(3\bar \O  +  \bar h
\ri)
\la{eq13a}\eea
with $h_S=S_r\,\O_{SS}, \,\bar h_S=S_r\,\bar \O, \, k_{\pp}=\bar{\pp}
\O_{\pp}$
and $\bar k_{\pp} =\bar \O$.
One cannot solve the eqs. (\ref{eq13a}) analytically and  the
numerical
solution  yields
\be
<S>=2.15, \;\;\;  <\pp>=-0.73, \;\;\;\;   c= 2 \times 10^{-15}
\la{eq14}\ee
where we have taken as an example $N_1=6, M_1=0, N_2=7, M_2=6$.
 This solution   corresponds to a minimum.

More general, a solution is a minimum if the Hessian $H= det \,V_{ab}
$ , where
$a, b$ denote derivatives with respect to real scalar fields and
$det$ the
determinant,  is positive definite. Let us define $p=Re\,S, \;
q=Im\,S, \;
\pp=\rho\, e^{i\theta}$ with $\rho >0 $.
 If we  only consider the contribution of the two gaugino
condensates, i.e.
there is no contribution of the chirial matter field $\pp$,  one has
$H=V_{pp}V_{qq} -V_{pq}^2$. Since  $q=Im\,S$ is an angular variable
there
will always be a minimum in the $q$ direction (i.e. $V_{qq} >0$).  In
fact, the
$q$ dependent part of the potential is $V(q)=A\, \cos \le[ q
(3/2b_1+3/2b_2)\ri]$ with $A$ independent of $q$ (for eq.(\ref{eq12})
we find
$A=2d_1d_2 [\; p \; (3/b_1+3/b_2) -1] \;e^{-p(3/2b_1 +3/2b_2)}$).
For
$V_q=0$ we necessarily have $V_{qp}=0$ and therefore the potential
has a
minimum in the $Re \,S, Im\,S$ plane if  and only  if $V_{pp} > 0$.
It is easy
to see that $h=0$ corresponds to $V_{pp} >0$ while $h  \neq 0$ to
$V_{pp} <0$.
In this case  $<S> \simeq 2.16$ and  we  see that the v.e.v. of the
dilaton  is
not very sensitive to the cancelation of the cosmological constant.
Using
eq.(\ref{eq14})  the difference on $<S>$ with or with out the
condition $V|=0$
is    $\Delta S =O(10^{-2})$.

Let us now consider  the Hessian including the contribution of $\pp$,
\begin{equation}
H =
 \left|
\begin{array}{cccc}
V_{pp} & 0  & V_{\rho p} & 0
\\
0 & V_{qq} & 0 & V_{ \theta q}
\\
V_{p \rho} & 0 & V_{\rho \rho} & 0
\\
0 & V_{q \theta} & 0 &  V_{\theta \theta}
\\
\end{array}
\right|
\ee
where  again  $V_{qp}=V_{q\rho}=V_{\theta p}=V_{\theta \rho}=0$ for
$V_q=V_\theta=0$. The Hessian
is positive definite if and only if $V_{pp}, V_{qq}, V_{\theta
\theta}, V_{
\rho \rho} > 0$ and
$H_1\equiv V_{pp}V_{qq} -V_{pq}^2= V_{pp}V_{qq} > 0, H_2\equiv
V_{pp}V_{ \rho
\rho} -V_{p \rho}^2 >0, H_3 \equiv V_{qq}V_{\theta
\theta}-V_{q\theta}^2>0$.
Since $\theta$ and $q$ are angular variables there will always be a
minimum in
those directions and in the $\theta, q$ plane, i.e. $H_3 >0$.   An
extremum
solution will be a minimum if and only if $V_{pp}, V_{\rho \rho}, H_2
> 0 $.
For the example given above we have checked numerically that $V_{pp}
> 0,
\;V_{\rho\rho}> 0$ and $H_2=V_{pp}V_{ \rho \rho} -V_{p \rho}^2> 0$
corresponding to a minimum. In \ci{stein} it was argued that a
potential with
$V|=V_S|=0$ always has a global minimum with $V<0$. However, their
 conclusion assumed that $V|=V_S|=0$  with  $W|=W_S|=0$ which  is not
our case
 since $W, W_S \neq 0$ at the minimum. In {\bf fig.1}  we show  $V_0$
as a
function of $S$.

We have thus shown that it is possible to cancel the cosmological
constant
using only the tree level sugra scalar potential. The dilaton field
has a
stable solution and its  v.e.v. does not differ much from the case
where
$\O_{ch}=0$ and  $V_{min} <0$. SUSY is  broken with  a large
hierarchy but
mainly due to the auxiliary field $k=\eta^{6} (T)\,G_\pp \sim
\O_{\pp}$  since
$G_T=0$ and $h=\eta^{6}(T) \,S_r\,G_s \sim \frac{\O\,\O_S}{\O_{SS}}$.
Note that
\be
\le| \frac{k}{h} \ri| \simeq	 \le| \frac{\O_{\pp}  \O_{SS}} { \O
\O_S} \ri| =
O(10^{3}) \gg 1
\ee
since $\O_{SS} \gg \O_S \sim \O $ and $\O_{\pp} \sim \O$.

A drop back is that many phenomenological  quantities   depend
strongly on how
SUSY is broken and  in this case it is broken via the term which we
now least
and was introduced with the only motivation of rendering $V|=0$.

\noindent
II) One gaugino condensate

We will now study the scalar potential  including loop corrections.
The dilaton
field is stabilized with a single gaugino condensate \ci{1gaug}.
The   Coleman-Weinberg  one-loop potential is \ci{V1}
\be
V_{1}=\frac{\Lambda^{4}}{64\pi^2}  Str \le[x+x^{2}
ln(\frac{x}{1+x})+ln(1+x)\ri]
\la{eq15}\ee
where $x=\frac{M^{2}}{\Lambda^{2}}$  and  $M^{2}$  are the (mass)$^2$
matrices
for the different particles. The leading contribution to
eq.(\ref{eq15}) comes
from the gaugino mass $m_g=
\frac{b_{0}}{6 \Lambda_c}e^{K/2}\,H$
with
\be
H\equiv \eta^{-6}\le[ F^{S}(K^{-1})^{S}_{S}(K_{S}
\O+\O_S)+K^i (K^{-1})^i_i (K_i \O +\O_i)-3\O \ri]
\la{eq16}\ee
 and $F_S=-(1+\frac{3S_r}{2b_0})$. Since
$x_{g}=\frac{m_g^{2}}{\Lambda^{2}_c}=O(10^{-2})$ we can write
eq.(\ref{eq15})
as
$V_{1}=-\frac{n_g}{16\pi^2}\Lambda_{c}^{2}m_g^{2}$ where $\Lambda_c$
is the
condensation scale and  $n_g$ is  the dimension of the hidden gauge
group
responsible for breaking SUSY.
  The scalar potential $V=V_{0}+V_{1}$ can then be easily written as
$
V=V_0 -\frac{n_g}{16\pi^2} \Lambda_c ^2 m^2_g$
\bea
V&=& \frac{1}{4}e^{K} \eta^{-12} \le[ |h|^2  +|k|^2 -3 |\O|^2 -
\delta |H|^2
\ri]
\no\\
V&=& \frac{1}{4}e^{K}\eta^{-12} [ |h|^2 (1-\de F^2_S) + |k|^2 (1-\de
|\pp|^2)-
3 |\O|^2 (1+3 \de)
\no\\
& \hspace{1cm}& -\de\le( F_S h [\pp\bar  k -3 \O] -3\pp \O  k  +
h.c.\ri) ]
\la{eq17}\eea
where we have defined  $\de \equiv\frac{n_g b_0^2}{144 \pi^2} \ll 1$
and  we
have specialized to $K_i=\bar \pp$.  We recover the tree level
potential by
setting $\de=0$. In the absence of chiral matter potential, i.e.
$k=0,$ the
dominant term
in  eq.(\ref{eq17}) is
\be
|h|^2 (1-\de F^2_S)=\le |d
e^{-3S/2b_0}\ri|^2\le(1+\frac{3S_r}{2b_0}\ri)\le(1-\de
\le[1+\frac{3S_r}{2b_0}\ri]^2\ri)
\la{eq18}\ee
where we have taken a single gaugino condensate $\O_0=d e^{-3S/2b_0}$
for which
$h=-\O_0 \le(1+\frac{3S_r}{2b_0}\ri)$. It is clear from
eqs.(\ref{eq17}) and
(\ref{eq18})  that there is  stable solution to $S$  with a single
gaugino
condensate unlike the previous case (tree level only).  Using this
approximation, the minimum is at
\be
S_r \simeq  \frac{2\,b_0}{3\, \delta ^{1/2}} = \frac{8\, \pi}{
\sqrt{n_g}}
\ee
and
\be
(1-\de F^2_S) \simeq - 2 \delta ^{-1/2} < 0.
\la{}\ee
For  $k \neq 0$  a  vanishing cosmological constant  requires
\be
|k|^2= - |h|^2 (1-\de F^2_S) + 3 |\O|^2
\ee
up to leading order. In this case SUSY is broken mainly due to the
auxiliary
field of the dilaton  since
\be
 \frac{|k|^2}{|h|^2} \simeq   -(1-\de F^2_S) = 2 \delta^{1/2}  \sim
10^{-2} .
\ee

Solving the exact eqs $V=V_S=V_\pp=0$ numerically, using  the ex. 1
in
\ci{1gaug}, i.e.  $SU(6)$ gauge group with $b_0=11/16\pi^2$,   the
v.e.v. of
$S$ and $\pp$ and the value of $c$ are
\be
S= 2.11,\;\;\;  \pp=0.74,\;\;\; c=2.9 \times 10^{-19}
\ee
corresponding to a stable solution, i.e. $V_{pp} > 0,\; V_{\rho \rho}
> 0$ and
$ H_2=V_{pp}V_{\rho \rho} -V_{p\rho}^2 > 0 $.  The difference between
the
v.e.v. of $S$ when  $c=\pp=0$ and $V < 0$ and when the condition
$V|=0$ is
imposed is small $\Delta \,S =O(10^{-3})$.

 In {\bf fig.2}  we show  $V$ as a function of $S$ and in {\bf fig.3}
we have
$V \;vs\; (\pp, S)$. We see that $V$ has two minimums in the $S$
direction.  In
the absence of  the chiral matter field contribution, i.e. $c=\pp=0$,
$V$ has
only one minimum and a similar behavior as the one shown in {\bf
fig.1}.
However, for $c\,\pp\neq 0$  there is a  local minimum with $V > 0 $.
Let as
call $S_0$ the global minimum, $S_2$ the local minimum and $S_1$ the
maximum in
between both minima.  In the region $S < S_0$ the potential $V_0$ and
$(- V_1)$
decrease exponentially fast and we have  $S_r
\O_S=\frac{3S_r}{2\,b_0}\O_0 \gg
c\,\pp$ and $x_g \ll 1$.
The dominant term in $V$ is $|h|^2 (1-\de F^2_S) + |k|^2 \sim |S_r
\O_S|^2
(1-\de F^2_S) + |c|^2$ and the minimum is at $S_0$.  For $S > S_1$
the tree
level potential decreases as $1/S$ since in this region $c\, \pp >
\O_S$ and
the dominant term in $V_0$ is then $e^{K} |c|^2 \sim |c|^2/S$.  On
the other
hand,  the one-loop potential $V_1$ decreases slightly (it becomes
more
negative) in the region $S_1 < S < S_2$ and therefore we have a local
minima at
$S_2$.  This behavior of $V=V_0+V_1 $ can be seen from
eq.(\ref{eq17}) since
the first two terms dominate and if   $S$ becomes larger then $|h|^2
(1-\de
F^2_S) + |k|^2 \sim
 -|c\,\pp|^2 \delta \le( \frac{3S_r}{2b_0}\ri)^2 +|c|^2 $ decreases.
For $S >
S_2$  the "log" terms in $V_1$ (cf. eq.(\ref{eq15})) start to
dominate,   $x_g
\gg  1$ and $V_1 \sim - \Lambda_c^4\, {\rm ln} \, x_g $ decreasing
exponentially fast.

Again, we have seen that it is  possible to cancel the cosmological
constant
while having $S$ at its minimum with a large hierarchy.

We will now consider the inflationary potential.  The dynamics of the
dilaton
field does not allow for an inflationary potential as long as $S$ is
still a
dynamically field.
A possible solution  is to have $S$ froze at its minimum and then
inflation
could be driven by other fields.  To be more general,   we will
investigate the
potential due to chiral matter fields  other then  $\pp$ which we
used to
cancel the cosmological constant\footnote{It would, nevertheless, be
interesting to study the phenomenology of  $\pp$.}. Clearly, a
potential to
inflate must be positive.  Is it then  possible to have $S$ stable
and $V > 0$?
 Can the inflaton potential dominate V ?

Let as take the potential $V$ to be independent of the inflaton
field, which we
will denote by $\Phi$. We assume $V$ to have the properties
$V|_{S_0}=V_S|_{S_0}=0$ at the global minima which we have jseen that
it is
indeed feasible.  The inflaton potential  $V_{inf}(S,\Phi)$ will in
general
depend on $S$. It is reasonable to assume that the $S$ dependent part
of
$V_{inf}$ can be factorized so that
\be
V_{inf} = A(S) P(\Phi).
\la{i1}\ee
For the  usual   inflaton potential in string scenarios one can have
$A=1/S
\sim e^{K}$
 or   $A=e^{-3 S/2b_0}/S \sim e^{K} W_0$  dependence on the dilaton
field.
The complete scalar potential is then
\be
V_f=V+V_{inf}.
\la{i2}\ee
Due to the inclusion of $V_{inf}$ the value of $S$ at the minimum
will now vary
and we will call it $S_1$
\be
V_{f,S}|_{S_1}=V_{S}|_{S_1} + V _{inf,S}|_{S_1}=0
\la{i3}\ee
which implies that
\be
P(\Phi) A_S|_{S_1} = -V_{S}|_{S_1} .
 \la{i4}\ee
Expanding $V$ and $V_S$ around $S_0$ and keeping the leading term
only, one has
\bea
V&=& \frac{1}{2} V_{SS}|_{S_0} (S - S_0 )^2 + ...
\no\\
V_{S}&=& V_{SS}|_{S_0} (S - S_0 ) + ...
\la{i5}\eea
since $V|_{S_0}=V_S|_{S_0}=0$ by hypothesis and $... $ correspond to
higher
derivative terms.

Let us take  $A_S = \alpha(S) A(S)$. For $A=1/S$ one has $\al = -1/S$
while for
$A=e^{-3 S/2b_0}/S $ one has  $\al = - 3/2b_0$ (keeping the leading
contribution only since  $ 3/2b_0 \gg 1$).  Using
eqs.(\ref{i3})-(\ref{i5}) we
can write the inflaton potential  $V_{inf} = A(S) P(\Phi)$ as
\bea
V_{inf}(S_1)&=&    \frac{1}{\al} A_S (S) P(\Phi)
\no\\
V_{inf}(S_1) &\simeq &-  \frac{1}{\al} V_{SS} |_{S_0} (S_1 - S_0 )
\la{i6}\eea
and the complete scalar potential is
\be
V_f (S_1)= V_{SS}|_{S_0}  (S_1 - S_0 ) \le( \frac{1}{2}   (S_1 - S_0
) -
\frac{1}{\al} \ri).
\la{i7}\ee
The first term in eq.(\ref{i7})  corresponds to the contribution from
$V$
while the second
comes from $V_{inf}$. Which terms dominates depends on $S_1-S_0$ and
$\al$.
For $\al  (S_1)=-1/S_1$ the potential is $V_{inf}=V_{SS} |_{S_0} (S_1
- S_0 )
\le( \frac{1}{2}   (S_1 - S_0 ) + S_1 \ri)$ and clearly the second
term, i.e.
$V_{inf}$  dominates. For $\al =-3/2b_0$ the potential is
$V_{inf}=V_{SS}|_{S_0}  (S_1 - S_0 ) \le( \frac{1}{2}   (S_1 - S_0 )
+ 2b_0/3
\ri)$ and $V_{inf}$ will dominate only if $    (S_1 - S_0 ) < 4b_0/3
$.  Note
that eqs.(\ref{i5}) and  (\ref{i6}) are only valid  if $S_1$ does not
differ
much from $S_0$ and it must be  calculated for  each  specific
example.
Furthermore,  the condition on  $S_1 - S_0 $  to be  small, i.e. that
$V_{inf}$
does not destabilize $S$,  imposes a constraint on the  magnitude of
$V_{inf}$
\ci{infaxel}.  Using  eqs.(\ref{i6})  and (\ref{i7}) the inflaton
potential
must satisfy
\be
\le| \frac{V_{inf}}{V_{SS}|_{S_0}} \ri| <   \frac{2 }{\alpha^2} .
\ee
This condition has been used to set an upper limit to the density
fluctuations
\ci{infaxel} and for specific examples it can be  consistent with the
COBE
observations \ci{cobe}.

To conclude, we have shown that in the context of sugra string models
the
cosmological constant  can be arranged to vanish   if  the
superpotential  has
a linear term.  The solution corresponds to a global minimum.  We
have also
shown that the introduction of a chiral potential $V_{inf}$ does not
necessarily destabilize the dilaton field  and the vacuum energy can
nevertheless be dominated by $V_{inf}$ allowing for the existence of
an
inflationary potential.

\noindent
{\bf Acknowledgment }

\noindent
It is a pleasure
 to thank  G. G. Ross for suggesting the problem and for many
enlightening
discussions and  comments.

\newpage

\noindent
{\bf Fig.1} {\small We show the potential  of two gaugino condensates
$V_0$ as
a function of $S$.}

\noindent
{\bf Fig.2} {\small We show  the potential of a single gaugino
condensate
$V=V_0+V_1$ as a function of $S$ for $c\,\pp \neq 0$.}

\noindent
{\bf Fig.3} {\small We show $V=V_0 + V_1$ as a function of $S$ and
$\pp$.}

\newpage

%%Begin InstantTeX Picture
\let\picnaturalsize=N
\def\picsize{4.8 in}
\def\picfilename{Grfv0s.eps}
%If you do not have the picture file add:
%\let\nopictures=Y
%to the beginning of the file.
\ifx\nopictures Y\else{\ifx\epsfloaded Y\else\input epsf \fi
\let\epsfloaded=Y
\centerline{\ifx\picnaturalsize N\epsfxsize \picsize\fi
\epsfbox{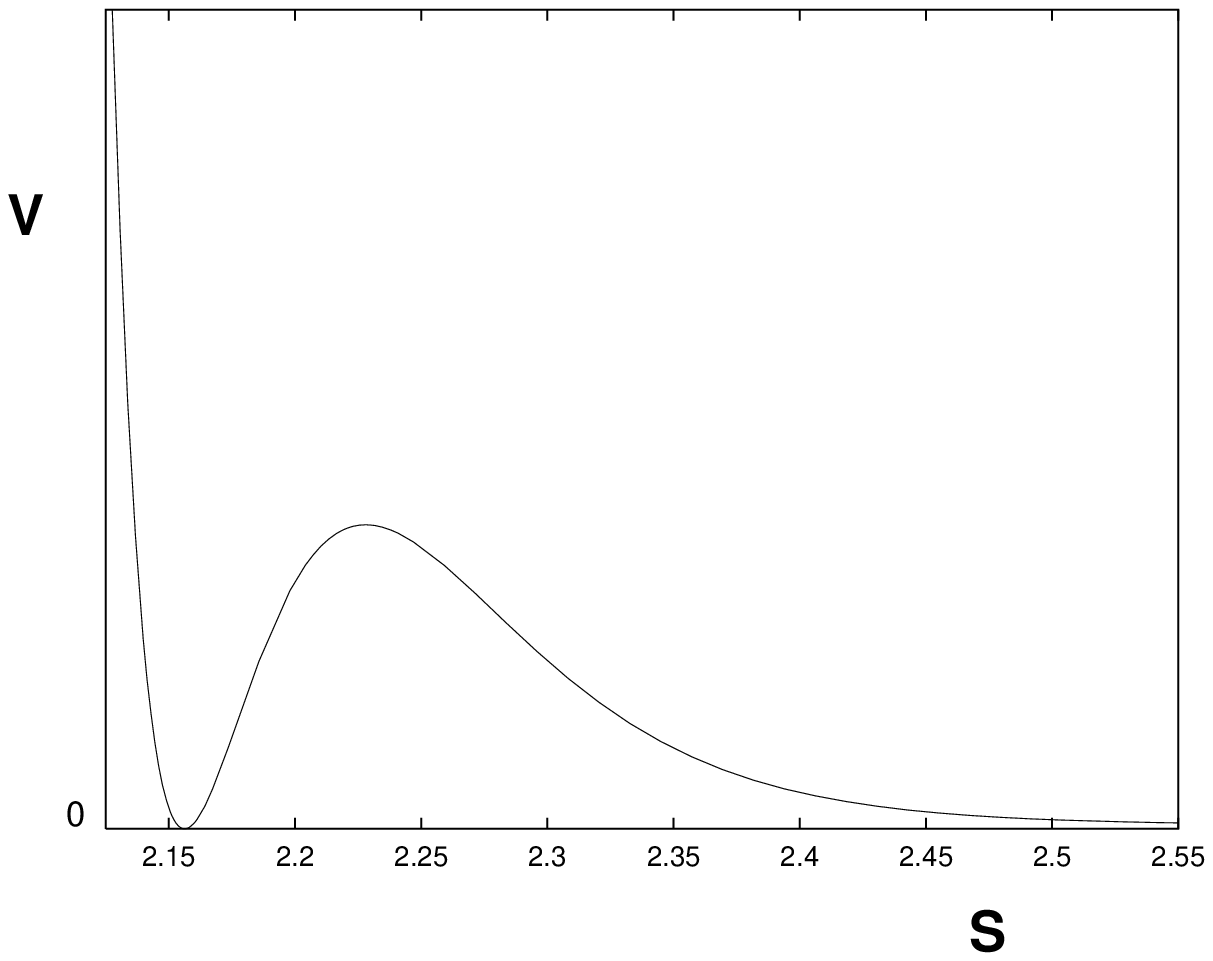}}}\fi
%%End InstantTeX Picture
 Fig.1 \\

\vspace{1cm}

%%Begin InstantTeX Picture
\let\picnaturalsize=N
\def\picsize{5.0in}
\def\picfilename{Grfvtots.eps}
%If you do not have the picture file add:
%\let\nopictures=Y
%to the beginning of the file.
\ifx\nopictures Y\else{\ifx\epsfloaded Y\else\input epsf \fi
\let\epsfloaded=Y
\centerline{\ifx\picnaturalsize N\epsfxsize \picsize\fi
\epsfbox{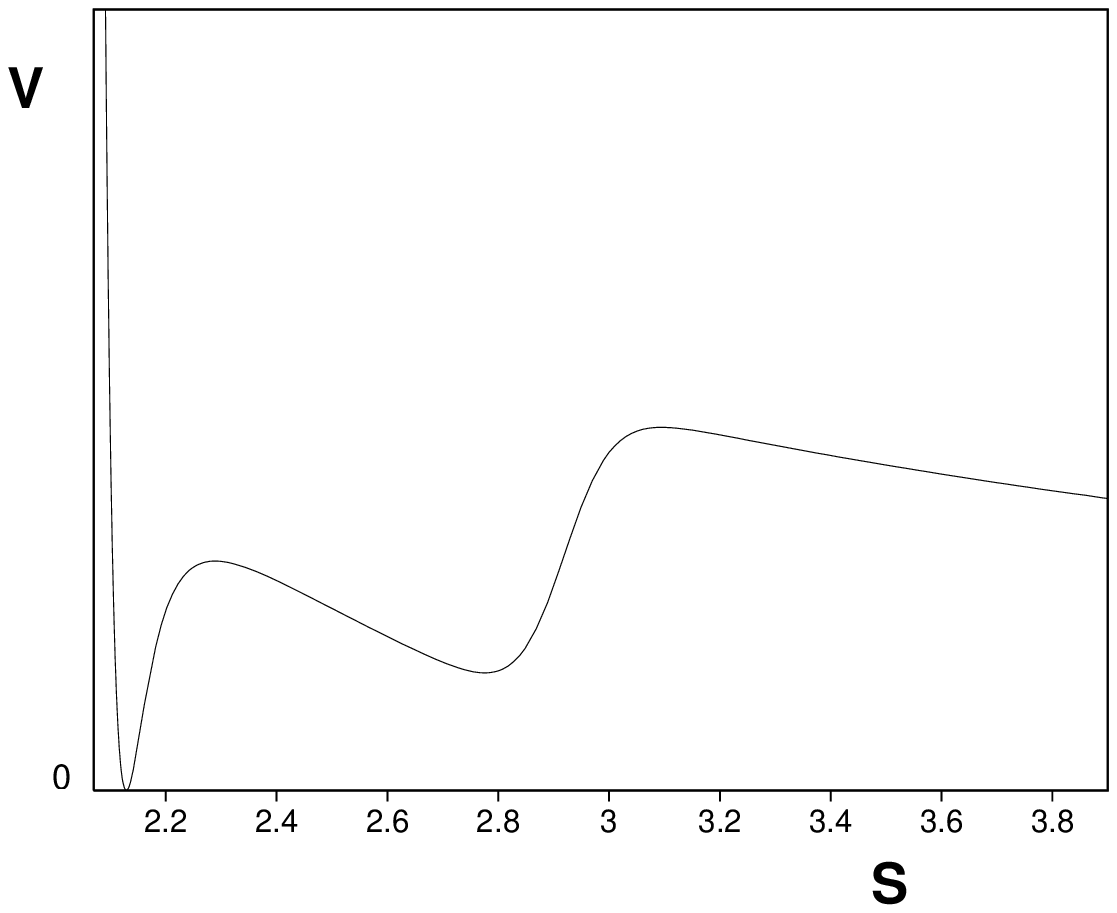}}}\fi
%%End InstantTeX Picture
Fig.2 \\

%\newpage

%%Begin InstantTeX Picture
\let\picnaturalsize=N
\def\picsize{5.0in}
\def\picfilename{Grfvtsx.eps}
%If you do not have the picture file add:
%\let\nopictures=Y
%to the beginning of the file.
\ifx\nopictures Y\else{\ifx\epsfloaded Y\else\input epsf \fi
\let\epsfloaded=Y
\centerline{\ifx\picnaturalsize N\epsfxsize \picsize\fi
\epsfbox{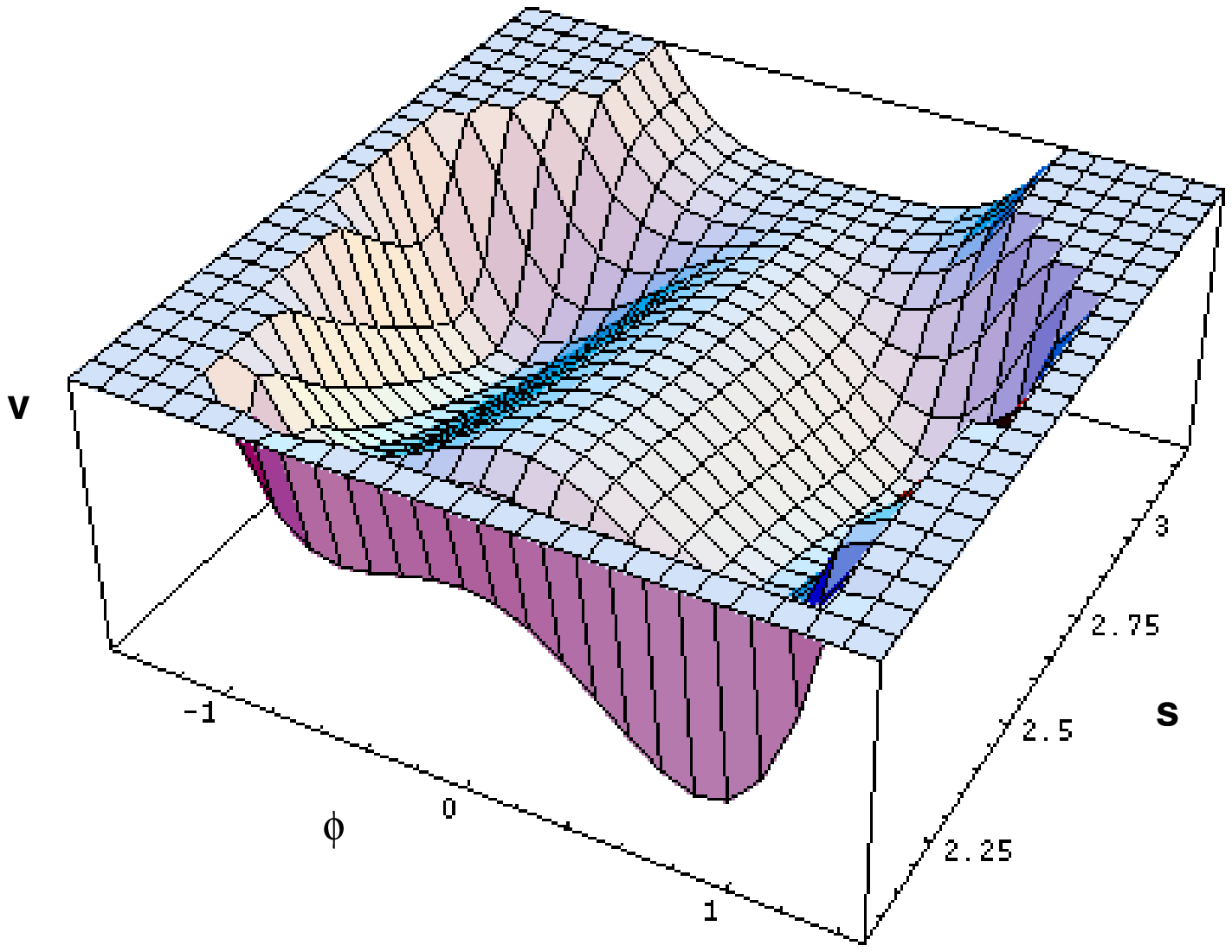}}}\fi
%%End InstantTeX Picture
Fig.3

\end{document}